\documentclass[aps,prd,twocolumn,preprintnumbers,superscriptaddress]{revtex4-1}
\usepackage{graphicx}
\usepackage{epstopdf}
\usepackage{amsmath}
\usepackage{amsfonts}
\usepackage{amssymb}
\usepackage{appendix}
\usepackage{enumerate}
\usepackage{natbib}
\usepackage{comment}
\usepackage{bbold}
\usepackage[shortlabels]{enumitem}
\usepackage{color}
\usepackage{slashed}
\usepackage{subfigure}
\usepackage{setspace}
\usepackage{footnote}
\usepackage{lipsum}
\usepackage{multirow}
\usepackage[colorlinks = true,
            linkcolor = blue,
            urlcolor  = blue,
            citecolor = blue,
            anchorcolor = blue]{hyperref}
\usepackage[capitalize]{cleveref}
\usepackage{braket}
\usepackage[compat=1.1.0]{tikz-feynman}
\usepackage{multirow}
\usepackage{physics}
\usepackage{feynmp-auto}
\usepackage[normalem]{ulem}
\usepackage{url}
\usepackage{units}
\usepackage[normalem]{ulem}

\newcommand{\be}{\begin{eqnarray}}
\newcommand{\ee}{\end{eqnarray}}
\newcommand{\ba} {\begin{equation}\begin{aligned}}
\newcommand{\ea} {\end{aligned}\end{equation}}
\newcommand{\bg} {\begin{equation}\begin{gathered}}
\newcommand{\eg} {\end{gathered}\end{equation}}

\newcommand{\beq}{\begin{equation}}
\newcommand{\eeq}{\end{equation}}

\usepackage{tikz,xcolor,hyperref}

\definecolor{lime}{HTML}{A6CE39}
\DeclareRobustCommand{\orcidicon}{\hspace{-1mm}
	\begin{tikzpicture}
		\draw[lime, fill=lime] (0,0) 
		circle [radius=0.12] 
		node[white] {{\fontfamily{qag}\selectfont \tiny \,ID}};
		\draw[white, fill=white] (-0.0525,0.095) 
		circle [radius=0.007];
	\end{tikzpicture}
	\hspace{-3mm}
}

\foreach \x in {A, ..., Z}{\expandafter\xdef\csname orcid\x\endcsname{\noexpand\href{https://orcid.org/\csname orcidauthor\x\endcsname}
		{\noexpand\orcidicon}}
}

\begin{document}

\title{Does the 220 PeV Event at KM3NeT Point to New Physics?}
	\author{Vedran~Brdar\orcidA{}}
	\email{vedran.brdar@okstate.edu}
	\affiliation{Department of Physics, Oklahoma State University, Stillwater, OK 74078, USA}
	\author{Dibya~S.~Chattopadhyay\orcidB{}}
	\email{dibya.chattopadhyay@okstate.edu}
	\affiliation{Department of Physics, Oklahoma State University, Stillwater, OK 74078, USA}

\begin{abstract}
The KM3NeT collaboration recently reported the observation of KM3-230213A, a neutrino event with an energy exceeding 100 PeV, more than an order of magnitude higher than the most energetic neutrino in IceCube's catalog. Given its longer data-taking period and larger effective area relative to KM3NeT, IceCube should have observed events around that energy. This tension has recently been quantified to lie between $2\sigma$ and $3.5\sigma$, depending on the neutrino source. A $\mathcal{O}(100)$ PeV neutrino detected at KM3NeT has traversed approximately $147$ km of rock and sea en route to the detector, whereas neutrinos arriving from the same location in the sky would have only traveled through about $14$ km of ice before reaching IceCube. We use this difference in propagation distance to address the tension between KM3NeT and IceCube. Specifically, we consider a scenario in which the source emits sterile neutrinos that partially convert to active neutrinos through oscillations. We scrutinize two such realizations, one where a new physics matter potential induces a resonance in sterile-to-active transitions and another one where off-diagonal neutrino non-standard interactions are employed. In both cases, sterile-to-active neutrino oscillations become relevant at length scales of $\sim100$  km, resulting in increased active neutrino flux near the KM3NeT detector, alleviating the tension between KM3NeT and IceCube. Overall, we propose the exciting possibility that neutrino telescopes may have started detecting new physics. 
\end{abstract}

\maketitle

\noindent
\textbf{Introduction.}
The field of high-energy neutrino astronomy kicked off in 2013 with the discovery of 
a number of neutrinos with energies between $100$ TeV and $1$ PeV by IceCube \cite{IceCube:2013low,IceCube:2014stg}. This count has steadily increased over the years  \cite{IceCube:2020wum}, including the observation of a Glashow resonance event \cite{IceCube:2021rpz}. IceCube has also identified  several high-energy neutrino sources, including the blazar TXS 0506+056 \cite{IceCube:2018cha,IceCube:2018dnn}, NGC 1068 active galaxy \cite{IceCube:2022der}, and the Milky Way \cite{IceCube:2023ame}. 
In parallel with IceCube, located at the South Pole, the ANTARES experiment was taking data in the Mediterranean Sea, using water instead of ice as the detection medium \cite{ANTARES:1999fhm,ANTARES:2011hfw}. ANTARES was succeeded by KM3NeT \cite{Margiotta:2014gza}, which consists of two detectors at different locations: ARCA and ORCA, where the former is specialized for high-energy astrophysical neutrino searches. It was recently reported \cite{KM3NeT:2025npi} (see also companion papers \cite{KM3NeT:2025vut,KM3NeT:2025bxl,KM3NeT:2025aps,KM3NeT:2025ccp,KM3NeT:2025mfl}) that ARCA detected a neutrino event on 13 February 2023, named KM3-230213A, with median neutrino energy of $E_\nu=220$ PeV. This sets the record for the highest-energy neutrino ever observed, exceeding by $\mathcal{O}(10)$ the highest-energy neutrino in IceCube's dataset. 

Comparing the effective areas and data-taking time across KM3NeT and IceCube, it is unexpected that KM3NeT would have been the first experiment observing $\mathcal{O}(100)$ PeV neutrinos. In fact, the observation of a single event at KM3NeT would 
point to the expectation of at least several events of similar energy reported by IceCube. The statistical significance of this tension was quantified in \cite{Li:2025tqf}, where the authors obtained results between $2\sigma$ and $3.5\sigma$, depending on whether the detected neutrino comes from a transient point source or has a diffuse origin. 
This~is also corroborated by KM3NeT collaboration which reported that a~$2.2\sigma$ upward fluctuation at KM3NeT is required in order to be consistent with the null observation at IceCube \cite{KM3NeT:2025npi}. 

\begin{figure}[b!]
    \centering
        \includegraphics[width=\linewidth]{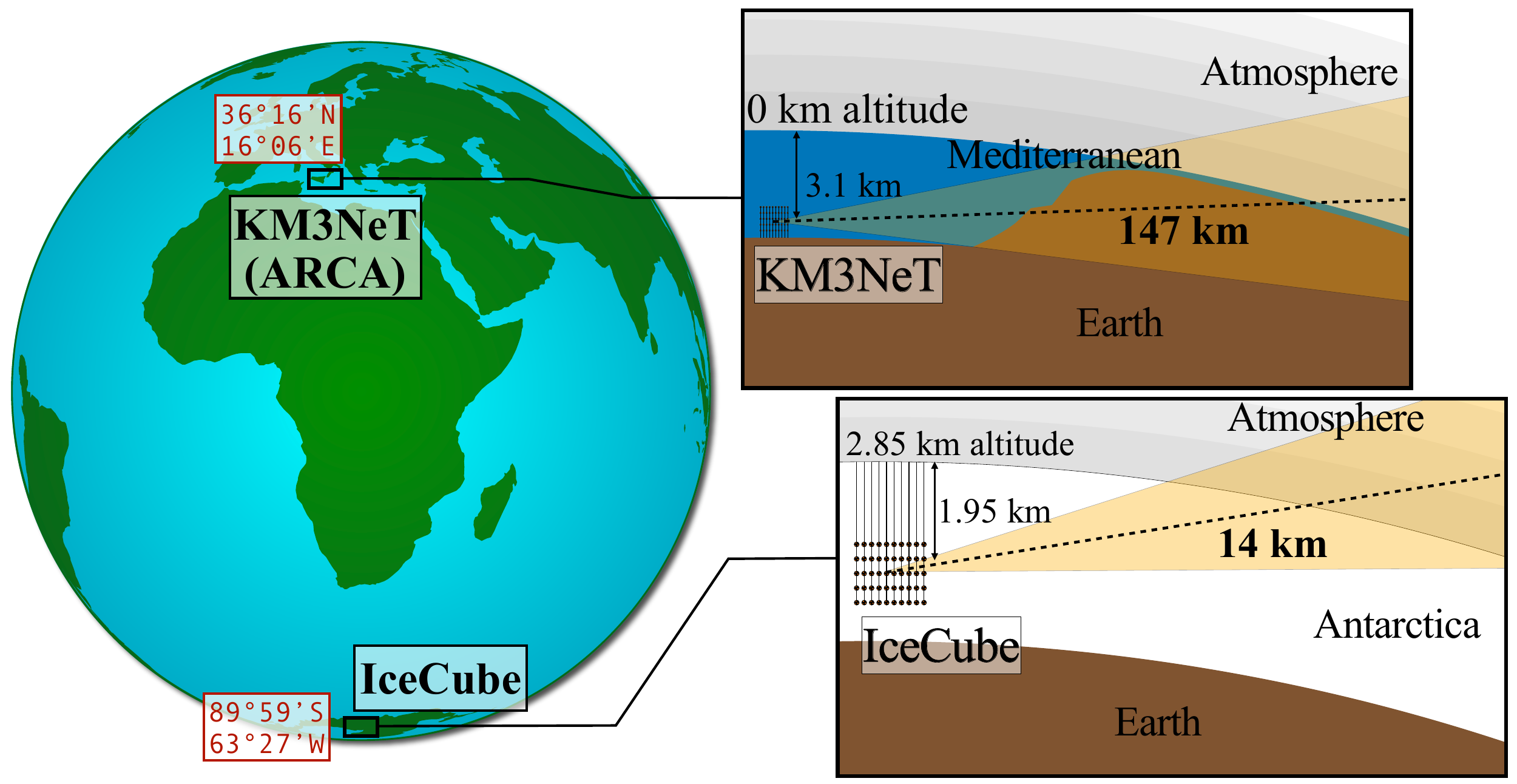}
    \caption{Topography of the vicinity of the KM3NeT and IceCube detectors in the nominal direction of an incoming $\mathcal{O}(100)$ PeV neutrino, as determined by the KM3NeT collaboration. We observe that the path through rock and sea for KM3NeT is an order of magnitude longer than the path through ice en route to the IceCube detector.}
    \label{fig:schematic}
\end{figure}

In this work, we discuss two scenarios that address this tension. What they have in common is a sterile neutrino. We assume that sterile neutrinos get copiously produced at a transient source and dominate in flux over active neutrinos. If that were not the case, active neutrino fluxes~at the KM3NeT and IceCube detectors would be similar, given the propagation distances in Earth which do not facilitate attenuation of neutrino flux for Standard Model (SM) cross sections. Even if there was a significant attenuation, this would reduce the flux at KM3NeT relative to IceCube, which would further strengthen the tension between the two experiments and is opposite from what we want to achieve.
High-energy sterile neutrinos could be produced directly in a transient source that is extremely dense. Sources such as choked GRBs or AGNs with a large column density of matter directly in the line of sight are promising candidates, where only weakly interacting particles, such as sterile neutrinos, could exit the production region with significant energy. Alternatively, dark sector interaction that stops active neutrinos could also be invoked, where only the sterile counterpart could traverse through a dense layer of dark sector and astrophysical distances.
The lack of multi-messenger observations~\cite{KM3NeT:2025bxl,Fang:2025nzg} in association to KM3-230213A further supports our premise.

The KM3NeT collaboration has very accurately determined the direction of the incoming $\mathcal{O}(100)$ PeV neutrino trajectory, with an uncertainty of $1.5^\circ$. 
The origin of KM3-230213A event corresponds to an incidence angle of $0.6^\circ$ at KM3NeT and $\sim 8^\circ$ above the horizon at IceCube. Using this, it is straightforward to determine the path length through rock and sea for KM3NeT, as well as the respective path length through ice en route to the IceCube detector for neutrinos coming from the same source. The topography is illustrated in \cref{fig:schematic}, where we show that the path through Earth for KM3NeT is an order of magnitude longer than that for IceCube. We employ this difference in propagation distance to alleviate the tension between IceCube and KM3NeT.
Note that, a single transient source is essential, as it uniquely fixes the Earth-traversed chord length and prevents IceCube from sampling comparable path-lengths for similar events happening at different coordinates of the sky. However, this particular source can be as long-lived as $\mathcal{O}(1)$ year, as the location of IceCube at the South Pole means that the path length traversed through Earth would constantly be $\mathcal{O}(10)$ km, whereas for KM3NeT, Earth’s rotation would allow it to have the optimal Line-of-Sight distance through Earth matter during some time of the day.
Namely, in what follows, we discuss two mechanisms where sterile neutrino flux partially transitions into active neutrino flux inside the Earth. Due to the longer propagation distance inside the Earth, the active neutrino flux at KM3NeT will be significantly higher compared to IceCube, compensating for the smaller effective area
, and thereby alleviating the tension.

\smallskip
\noindent
\textbf{Neutrino production through matter-induced resonance.}
When the vacuum mixing between active~and sterile neutrinos is very small,
matter can resonantly~amplify sterile-to-active transitions. The MSW resonance~\cite{Wolfenstein:1977ue,Mikheyev:1985zog,Mikheev:1986wj}  is realized when the matter potential difference for the involved neutrino species, $V$, equals $[\Delta m^2/(2 E_\nu)] \cos 2\theta$, where $\Delta m^2$ and $\theta$ are the mass-squared difference and vacuum mixing angle, respectively. Taking the values from the SM, matter potential reads $V \equiv V_\text{CC} = \sqrt{2} G_F n_e \simeq 3.7\times 10^{-23}$ GeV for matter with a density of $1\, \text{g}/\text{cm}^3$. The resonant oscillation length then reads $L_{\text{res}}\simeq \pi/(V \sin 2\theta)\approx 5\times10^3 \,\theta^{-1}$ km in the limit of small $\theta$. For keV-scale sterile neutrino, which will turn out to be required in our analysis, maximum mixing angle value of $\theta^2\simeq 10^{-3}$ is allowed in light of terrestrial experiments \cite{MINOS:2017cae,Bolton:2019pcu}.
Such values are in tension with cosmological constraints, but those can be alleviated \cite{Dasgupta:2013zpn,Cherry:2016jol,Chu:2018gxk,Farzan:2019yvo,Cline:2019seo,Gelmini:2019esj,Barbieri:2025moq}.
This yields $L_{\text{res}}\simeq 10^5$ km or longer, and for $L\approx 150$ km, which corresponds to KM3NeT, the oscillation probability would then not surpass $L^2/L_{\text{res}}^2 \approx 10^{-6}$.

This shortcoming can be addressed by considering a large sterile neutrino matter potential, a scenario that has been widely explored in the literature \cite{Pospelov:2011ha,Pospelov:2012gm,Kopp:2014fha,Denton:2018dqq,Berryman:2018jxt}. Specifically, we will focus on the realization introduced in \cite{Pospelov:2011ha}, where a light vector boson, $V^\mu$, associated with a spontaneously broken $U(1)_B$ symmetry, interacts with a sterile neutrino, $\nu_s$, and quarks \cite{Pospelov:2011ha} 
\begin{align}
\mathcal{L}\supset g_b' \bar{\nu}_s \slashed{V} \nu_s +(g_b/3)\sum_q  \bar{q} \slashed{V} q + \text{h.c.}\,.
\end{align}
Here, $g_b'$ is a product of the gauge coupling $g_b$ and a sterile neutrino charge under a new symmetry group. Since we are interested in the matter potential associated with scattering at momentum transfer $Q^2=0$, it is convenient to represent the above terms in the effective field theory framework \cite{Pospelov:2011ha,Kopp:2014fha}  
\begin{align}
\mathcal{L}_\text{eff}\supset \frac{G_B}{2} \big[\bar{\nu}_s \gamma_\mu (1-\gamma_5) \nu_s] [\bar{p}\gamma^\mu p+\bar{n} \gamma^\mu n\big]\,.
\label{eq:lowE}
\end{align}
This expression is obtained after integrating out $V^\mu$ (note that the mass of the introduced vector boson, $m_V$, always satisfies $m_V^2\gg Q^2 =0$). In \cref{eq:lowE}, $p$ and $n$ represent protons and neutrons, and $G_B=g_b g_b'/m_V^2$. The sterile neutrino matter potential can then be written as $V_s=G_B n_\text{nuc}$ \cite{Pospelov:2011ha,Kopp:2014fha},
where $n_\text{nuc}=n_p+n_n$ is the nucleon number density in matter. For $V_s\gg G_F n_e$, the difference between the sterile and active neutrino matter potential, $V$, is approximately equal to $V_s$. We parameterize $V_s$ using the dimensionless quantity 
$\epsilon_{ss}=G_B/(\sqrt{2} G_F)$, which represents the ratio of the new baryonic interaction strength to Fermi's interaction strength. In the literature \cite{Pospelov:2011ha,Harnik:2012ni,Pospelov:2012gm}, $\epsilon_{ss}$ values up to $\mathcal{O}(10^2-10^3)$ have been explored. Upon investigating the effects of $\epsilon_{ss}$ on neutrino oscillations, we find that values $|\epsilon_{ss}| < 10^3$ typically lead to deviations smaller than a few percent for atmospheric neutrinos. Note that such an interaction strength precisely addresses the $L^2/L_{\text{res}}^2\ll 1$ problem for KM3NeT and makes the propagation length in Earth comparable to the resonant oscillation length. We should stress that limits on $\epsilon_{ss}$ become more severe if a sterile neutrino mixes with several active neutrinos; constraints arising from solar neutrino data in a scenario where a sterile neutrino mixes with an electron neutrino are particularly restrictive \cite{Kopp:2014fha}. However, in this work, we consider sterile neutrino mixing only with a muon neutrino. The generation of muon neutrino flux via oscillations is crucial, as muon neutrinos can produce muon tracks at KM3NeT through SM charged current interactions, explaining the recent observation. For $E_\nu=220$ PeV and propagation in water/rock ($\rho\approx 1-2.2 \, \text{g}/\text{cm}^3$, $n_\text{nuc}\approx  \rho/m_p$), MSW resonance occurs for $\sqrt{\Delta m^2}\approx m_s \simeq 2\times 10^{-1} \sqrt{|\epsilon_{ss}|}$ keV, where $m_s$ is the sterile neutrino mass. 
For $-\epsilon_{ss}\simeq\mathcal{O}(100)$, a keV-scale sterile neutrino will undergo resonance in the Earth en route to the KM3NeT detector. Propagation of sterile neutrinos in the direction corresponding to the detected event at KM3NeT involves passage through both rock and water (see \cref{fig:schematic}); we take the former (latter) to be $100$ km ($47$ km) long, with rock (water) density of $2.2~\rm{g}/\rm{cm}^3$ ($1~\rm{g}/\rm{cm}^3$). We assume constant density matter, with an average density of $1.82~\rm{g}/\rm{cm}^3$. A more detailed treatment with varying densities would only change this picture marginally.
To this end, we solve the Schr{\"o}dinger equation which is governed by the following Hamiltonian in the ($\nu_\mu$, $\nu_s$) flavor basis
\begin{align}
	\hspace{-3pt}
H= \! \begin{pmatrix} c_\theta & s_\theta \\ -s_\theta & c_\theta \end{pmatrix}
\begin{pmatrix} 0 & 0 \\ 0 & \frac{m_s^2}{2 E_\nu} \end{pmatrix}
\begin{pmatrix} c_\theta & -s_\theta \\ s_\theta & c_\theta \end{pmatrix} \! + \!
 \begin{pmatrix} V_{\text{NC}} & 0 \\ 0 &  V_s \end{pmatrix} ,
\end{align}
where we have abbreviated  $c_\theta=\cos \theta$ and $s_\theta=\sin \theta$.
The term $V_{\text{NC}}=-G_F n_n/\sqrt{2} \approx -V_{\text{CC}}/2$ is the SM neutral current matter potential which, while included in our analysis, will not impact results due to its smallness with respect to $V_s=\sqrt{2} G_F \epsilon_{ss} (n_n+n_p)$. Further, we simplify $n_p \approx n_n$, i.e., $V_s \approx 2 \epsilon_{ss} V_{\rm CC}$.
The resonance  occurs in the neutrino sector for negative values of $\epsilon_{ss}$ and this is a scenario that we will adopt (for positive values of $\epsilon_{ss}$ resonance would be observed in the antineutrino sector). 

\begin{figure}[t!]
    \centering
        \includegraphics[width=0.98\linewidth]{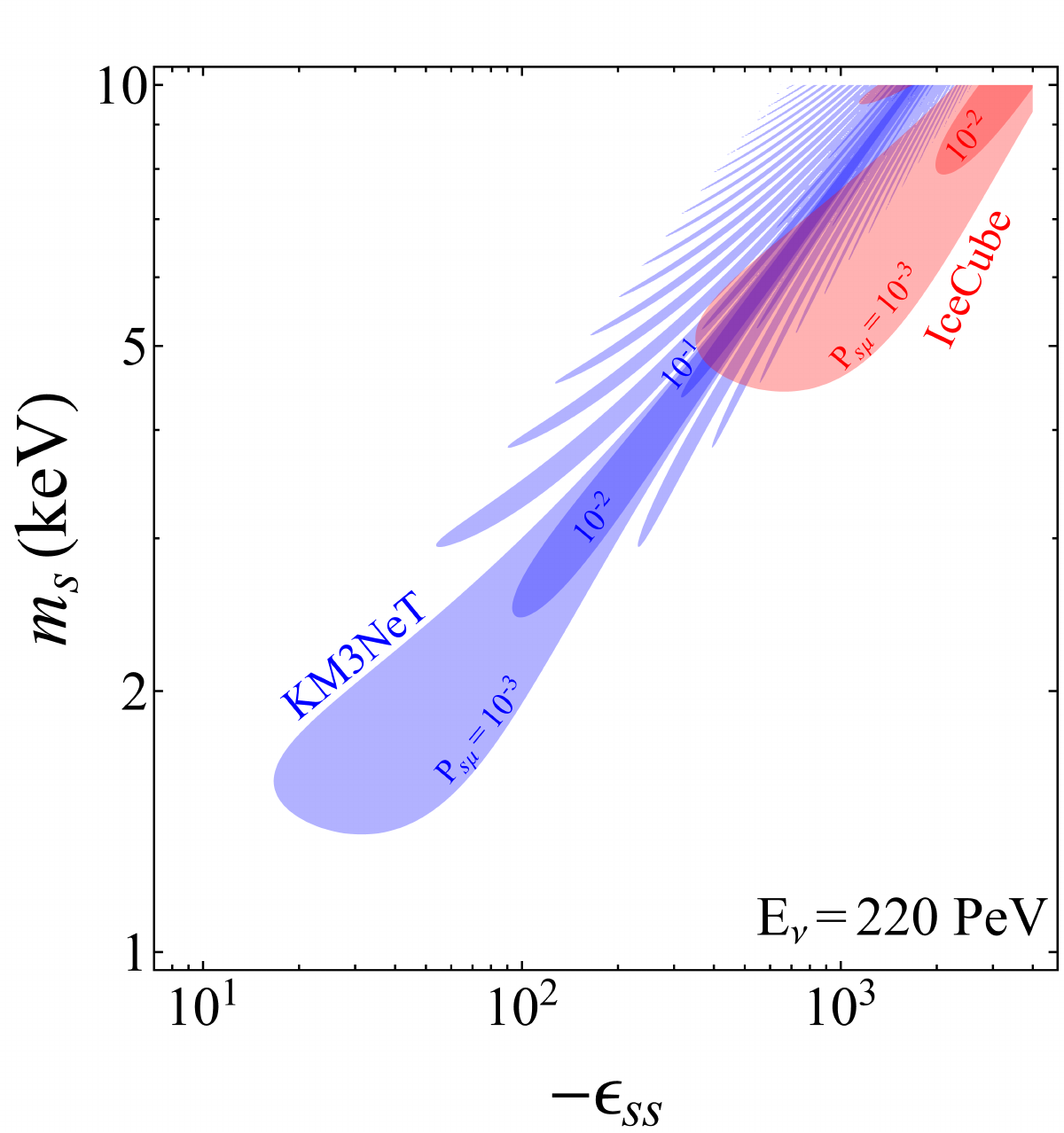}
    \caption{Sterile-to-active neutrino oscillation probability, $P_{s\mu}$, is shown in the parameter space of the sterile neutrino mass $m_s$ and interaction strength 
$-\epsilon_{ss}$ for both KM3NeT and IceCube. The vacuum mixing angle is fixed to $\theta=10^{-2}$. Deeper shades of blue and red correspond to higher oscillation probabilities at KM3NeT and IceCube, respectively.}
    \label{fig:baryonic1}
\end{figure}

\begin{figure}[t!]
    \centering
        \includegraphics[width=\linewidth]{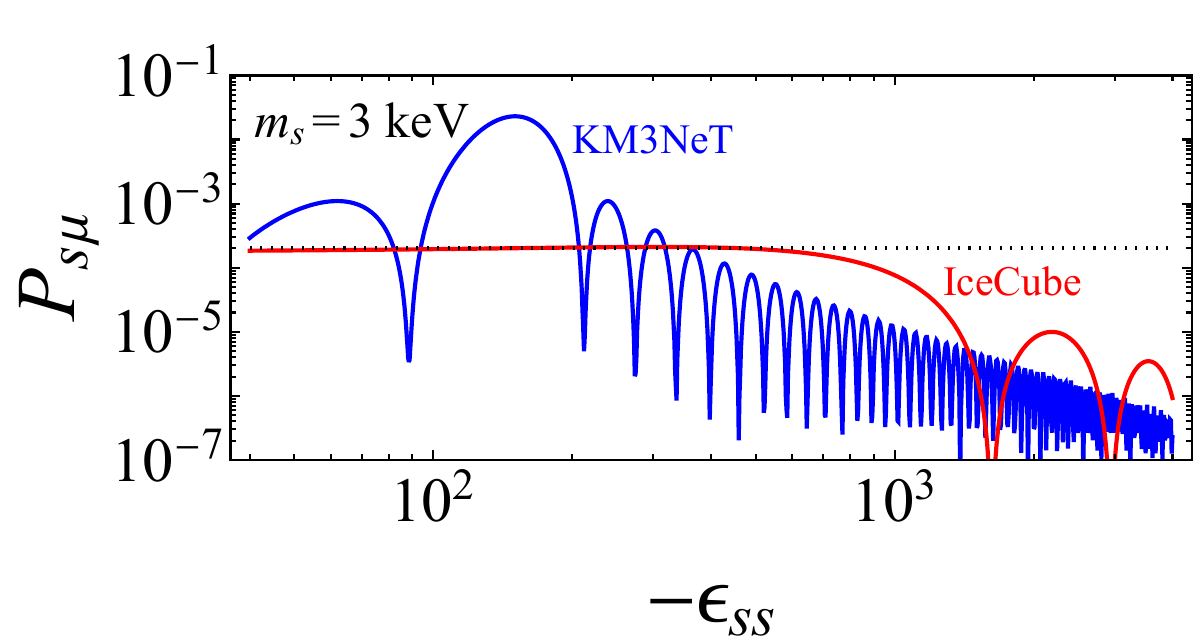}
    \caption{Sterile-to-active neutrino oscillation probability, $P_{s\mu}$, is shown as a function of interaction strength $-\epsilon_{ss}$ for both KM3NeT and IceCube, for a sterile neutrino mass of $m_s=3$ keV and a vacuum mixing angle $\theta=10^{-2}$.}
    \label{fig:baryonic2}
\end{figure}

The results of flavor evolution in the Earth for both IceCube and KM3NeT are shown in \cref{fig:baryonic1,fig:baryonic2} (for more details regarding the energy dependence, see the Supplemental Material~\cite{supplemental}); in both figures, the vacuum mixing angle is fixed to $\theta=10^{-2}$ and we present oscillation probabilities after traversing 147 km and 14 km for KM3NeT and IceCube, respectively. In \cref{fig:baryonic1}, we show sterile-to-active neutrino oscillation probability, $P_{s\mu}$, in the parameter space of sterile neutrino mass (scanned between $1$ and $10$ keV, values for which a resonance is expected) and $-\epsilon_{ss}$. For KM3NeT, sizable values of $P_{s\mu}$ already appear for $-\epsilon_{ss}\gtrsim 30$. In the region between $-\epsilon_{ss}\simeq 100$ and $-\epsilon_{ss}\simeq 10^3$, $P_{s\mu}$ for KM3NeT attains values of $P_{s\mu}\gtrsim 10^{-2}$, implying a relatively large muon neutrino flux generated in Earth. At around $-\epsilon_{ss}\sim 10^3$, the oscillation probability for IceCube reaches $P_{s\mu}\simeq 10^{-3}$, remaining smaller by $1$--$2$ orders of magnitude compared to that of KM3NeT. This is because $P_{s\mu}$ depends on $\sin^2(L/L_{\text{res}})$, and the neutrino propagation length in Earth (along the direction of the 220 PeV neutrino detected by KM3NeT) is about 10 times shorter at IceCube. Therefore, at resonance, $P_{s\mu}$ is always expected to be at least two orders of magnitude larger for KM3NeT. The figure can be extended to larger values of $-\epsilon_{ss}$ that would require heavier sterile neutrinos to induce a resonance. However, increasing the interaction strength would create tension with the data \cite{Pospelov:2011ha,Harnik:2012ni,Pospelov:2012gm}.
As argued in \cite{Harnik:2012ni},  for a leptophobic $U(1)_B$ gauge boson that couples only to baryon number, many bounds are not applicable or they weaken. For example, with the leading production mechanism being through radiation off nuclei, stellar cooling bounds would be suppressed~\cite{Harnik:2012ni}.
Further, supernova bounds for this scenario go only up-to light vector boson masses of $\sim0.2$~GeV~\cite{Rrapaj:2015wgs}. Some of the most stringent constraints on such a model arise from anomaly cancellation~\cite{Dror:2017ehi}, with constraints for the vector boson coupling to baryons at $g_b \lesssim 1.85 \times 10^{-3}$, for a benchmark mass of $\sim 0.5$~GeV. The coupling of the vector boson with the sterile sector can be large. If we allow a maximum coupling of $g_b' \approx 1$ (remaining well within the perturbative limit), we obtain a maximum allowed value of $|\epsilon_{ss}| \lesssim 450$. Note that this completely rules out the red region in~\cref{fig:baryonic1}, preferred for a larger $\nu_s \to \nu_\mu$ oscillation probability at IceCube.

 In \cref{fig:baryonic2}, we show $P_{s\mu}$ as a function of $-\epsilon_{ss}$ for both KM3NeT and IceCube and for a fixed sterile neutrino mass of $m_s=3$ keV. The resonance is obtained for $-\epsilon_{ss}\simeq 150$, a parameter point where $P_{s\mu}$ for KM3NeT is two orders of magnitude larger than for IceCube. This difference in oscillation probability, i.e., in the generated muon neutrino flux, alleviates the tension between the two experiments by compensating for KM3NeT's smaller effective area and shorter data-taking period. 
In \cref{fig:baryonic2}, we also show the dotted line, which represents the probability 
for sterile-to-active oscillations in vacuum, i.e. the muon neutrino flux generated between the source and the Earth relative to the sterile neutrino flux.
The muon neutrino flux generated in vacuum is subdominant compared to the one generated through resonance, as required, since otherwise the event rate would be dominated by neutrinos produced outside the Earth, which would not explain the tension.

Before closing this section, we should also stress that the air density at the KM3NeT location (sea level) is 20\% larger than the one at IceCube \cite{atmos2,atmos1}. Therefore, in principle, a resonance in the atmosphere can be achieved at KM3NeT and avoided at IceCube (see \cite{Sponsler:2024iej} for resonant neutrino flavor conversion in the atmosphere). However, since rock and water have densities three orders of magnitude larger than air at sea level, the oscillation length at resonance in air would be significantly longer, and we do not anticipate strong oscillation effects in such a case.

\smallskip
\noindent
\textbf{Neutrino production via non-standard interactions.}
In what follows, we present an oscillatory realization that alleviates the tension between IceCube and KM3NeT without involving resonance. We consider the following Hamiltonian in the ($\nu_\mu,\nu_s$) flavor basis
\begin{align}
H=&\begin{pmatrix} c_\theta & s_\theta \\ -s_\theta & c_\theta \end{pmatrix}
\begin{pmatrix} 0 & 0 \\ 0 & {m_s^2}/{(2 E_\nu)} \end{pmatrix}
\begin{pmatrix} c_\theta & -s_\theta \\ s_\theta & c_\theta \end{pmatrix} \nonumber \\ &+
\begin{pmatrix} V_{\text{NC}} & \epsilon_{\mu s} V_{\text{CC}} \\ \epsilon_{\mu s} V_{\text{CC}} & 0 \end{pmatrix}  \,,
\label{eq:NSI}
\end{align} 
where we now turn on the off-diagonal element in the matter potential. This type of lepton flavor violating term has been constrained in the literature for the active neutrino sector \cite{Farzan:2017xzy,Esteban:2018ppq,Proceedings:2019qno,Coloma:2023ixt}. Overall, $\mathcal{O}(10^{-1}-1)$ values of $\epsilon_{e\mu}$ remain unconstrained \cite{Esteban:2018ppq,Coloma:2023ixt}, and such large non-standard interactions can be generated in UV-complete scenarios \cite{Farzan:2015hkd,Farzan:2016wym,Farzan:2019xor}. As can be seen from \cref{eq:NSI}, we have an off-diagonal element mixing the muon and sterile neutrino. This term, parameterized by $\epsilon_{\mu s}$, arises from the following effective Lagrangian
\begin{align}
\mathcal{L}\supset -2\sqrt{2} G_F \epsilon_{\mu s}^f \,(\bar{\nu}_s \gamma^\mu P_L \nu_\mu) (\bar{f} \gamma_\mu f)\,,
\label{eq:lagrangian}
\end{align} 
where $f$ represents SM charged fermions. We will not focus on model-building aspects of this interaction and will instead examine its phenomenological implications, primarily in explaining the tension between KM3NeT and IceCube. Note that the SM fermions that neutrinos interact with in \cref{eq:lagrangian} do not necessarily need to be protons and neutrons, as in our previously discussed case of baryonic interaction; interaction with electrons is viable as well. The $\epsilon_{\mu s}$ coefficient can generally be a complex number, but for simplicity, we take it to be real and positive. 
\begin{figure}[t!]
    \centering
    \includegraphics[width=\linewidth]{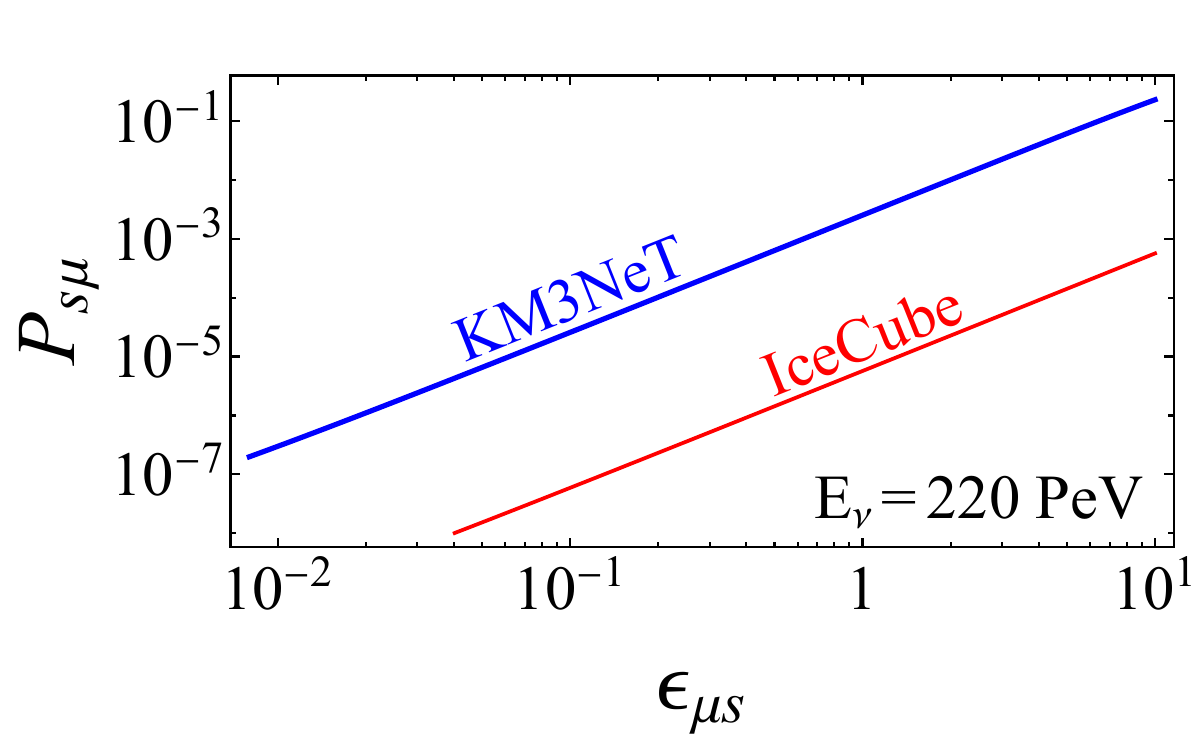}
    \caption{The sterile-to-active neutrino oscillation probability, $P_{s\mu}$, in the scenario with off-diagonal non-standard neutrino interactions is shown for both KM3NeT and IceCube for \mbox{$m_s=500$~eV}.}
    \label{fig:sterileNSI}
\end{figure}
In our scenario, a sterile neutrino with $E_\nu = 220$ PeV would mix with the active neutrino sector due to matter effects parameterized by $\epsilon_{\mu s}$. The effective mixing angle in matter in the limit of a tiny vacuum mixing angle $\theta$, large $m_s$, and $V_{\rm NC} = -V_{\rm CC}/2$ takes the form  
\begin{align}
\theta_m = \frac{1}{2} \tan^{-1} \left( \frac{4 \epsilon_{\mu s}  E_\nu V_{\text{CC}}}
{m_s^2+ E_\nu V_{\text{CC}}} \right) \,.
\end{align}
In the limit of $V_\text{CC}\gg m_s^2/E_\nu$, this further reduces to $\theta_m = (1/2) \tan^{-1} (4\epsilon_{\mu s})$. The effective mass squared difference in matter for $\theta \approx 0$ and $V_{\rm NC} = -V_{\rm CC}/2$ reads
\begin{align}
m_\text{eff}^2 = \sqrt{\left(E_\nu V_{\text{CC}} + m_s^2 \right)^2 
+ 16 \epsilon_{\mu s}^2 E_\nu^2 V_{\text{CC}}^2
}\,\,.
\end{align}
Using these results, the conversion probability from sterile to muon neutrino is given by  
\begin{align}
P_{s \mu} = \sin^2 (2\theta_m) \sin^2 \left[{m_\text{eff}^2 L}/{(4 E_\nu)}  \right] \, \,,
\label{eq:P}
\end{align}
where $L$ is neutrino propagation length in Earth. In \cref{fig:sterileNSI}, we show the conversion probabilities for KM3NeT and IceCube as a function of $\epsilon_{\mu s}$ for $m_s = 500 \text{ eV}$ and $\theta=10^{-4}$. We have scrutinized the impact of the considered interaction on $P_{ee}$, $P_{\mu e}$, and $P_{\mu\mu}$ at low energies and found that $\epsilon_{\mu s}$ values up to $\mathcal{O}(1-10)$ have negligible effect on neutrino oscillations for $m_s \gtrsim 100$ eV. Hence, in the figure, we present oscillation probabilities for $\epsilon_{\mu s}$ values up to $10$. The sterile-to-active oscillation probability for KM3NeT is larger than that for IceCube by roughly two orders of magnitude for all presented values of $\epsilon_{\mu s}$. This can be understood from the oscillation formula in \cref{eq:P}, which features an $L^2$ dependence when ${m_\text{eff}^2 L}/{(4 E_\nu)} \lesssim \mathcal{O}(1)$, which is satisfied. Thus, the difference in the propagation length between KM3NeT and IceCube leads to the observed difference in the value of $P_{s\mu}$. While the ratio of $P_{s\mu}$ for KM3NeT and IceCube does not depend on $\epsilon_{\mu s}$, it should be noted that solutions with larger $\epsilon_{\mu s}$ are more favorable, as they lead to a larger generated active neutrino flux at KM3NeT, and thus require smaller incoming flux of sterile neutrinos. As in our previous scenario with resonance, the larger muon neutrino flux generated en route to KM3NeT alleviates the tension arising from the non-observation of $\mathcal{O}(100)$ PeV neutrinos in IceCube.  For the energy dependence of the non-resonant scenario, see the Supplemental Material.

Note that, both the scenarios presented in our letter ensure that only sterile neutrinos at a particular energy range (and beyond) oscillate into active neutrinos in the presence of matter, so it is allowed for this individual transient event to have a much broader energy range in the incoming sterile flux. Namely, $\mathcal{O}(100)$ PeV neutrinos can oscillate into active neutrinos for the resonant solution. Whereas, for the non-standard interaction solution, for $E_\nu V_{\text{CC}} \gg m_s^2$ the sterile to active conversion probability would be enhanced significantly in the presence of matter effects; higher energies would be suppressed by an expected power law drop in the sterile neutrino flux.

In our study, we consider a transient sterile neutrino source; the required luminosity to observe a single event at KM3NeT depends on the emission duration from the source, the conversion probability $P_{s\mu}$ at Earth, and the source distance. For a purely sterile transient source, at a distance of $100$~Mpc, lasting $\sim \mathcal{O}(10^3)$~s during the optimal observation window, the required time-averaged sterile neutrino luminosity in the  $\mathcal{O}(150 - 300)$~PeV band would need to be $\sim \mathcal{O}(10^{50}-10^{51})$ erg/s.
Larger values of $m_s$ and $\epsilon_{ss}$ (or $\epsilon_{\mu s}$), satisfying either the resonance condition or those corresponding to the non-standard interaction scenario, can lead to an enhanced $P_{s\mu}$ which can reduce the required luminosity by a factor of $\sim\mathcal{O}(10-100)$.

\noindent
\textbf{Summary and Outlook.}
KM3NeT has recently observed the highest-energy neutrino event exceeding $\mathcal{O}(100)$ PeV, and the lack of observation of neutrinos at such energy at IceCube, which has a larger effective area and a longer data-taking time, is unexpected. 
We present for the first time a new physics solution to this tension that has recently been quantified to be between $2\sigma$ and $3.5\sigma$. We consider a scenario where most of the flux from a point source in the nominal direction of a detected $\mathcal{O}(100)$ PeV neutrino, arrives at Earth as sterile neutrinos. Muon neutrinos are then generated via sterile-to-active oscillations in Earth en route to the KM3NeT and IceCube detectors. Given the direction of the incoming $\mathcal{O}(100)$ PeV neutrino trajectory, the path through Earth for KM3NeT is an order of magnitude longer than for IceCube. This discrepancy in propagation length yields a higher flux of generated muon neutrinos at KM3NeT, which compensates for KM3NeT's deficiency in effective area and data-taking time, successfully explaining why IceCube has not detected neutrinos at such high energies. We have presented two scenarios that feature strong oscillations between sterile and muon neutrinos: one in which oscillations are resonantly amplified in matter through the sterile neutrino interaction with baryons, and another featuring off-diagonal non-standard interactions. We would like to stress that oscillations may not be the only option for successfully alleviating the tension between KM3NeT and IceCube. In particular, scattering in Earth may be considered, as the likelihood for interaction is proportional to the propagation length in Earth, setting the stage for resolving the tension. However, we have not found a successful solution to the tension when a sterile neutrino produces a muon neutrino via scattering, even in the framework where very large Glashow-like cross sections at $\mathcal{O}(100)$ PeV energies are employed. This is due to the excess of \emph{sterile} neutrino scattering events within the fiducial volume of IceCube and KM3NeT. We leave further investigation of the scattering method for future work, where we will focus on extended hidden sectors.

\noindent
\textbf{Acknowledgments.}
The work of VB is supported by the United States Department of Energy Grant No.
DE-SC0025477.

\bibliographystyle{JHEP}
\bibliography{refs}

\newpage

\appendix

\clearpage
\newpage
\widetext

\begin{center}
	\vspace{0.08in}
	
	\textbf{\large{Does the 220 PeV Event at KM3NeT Point to New Physics?}}
	
	\vspace{0.1in}
	\emph{\large{Supplemental Material}} \\[6pt]
	{Vedran Brdar, Dibya S. Chattopadhyay}
\end{center}

\setcounter{equation}{0}
\setcounter{figure}{0}
\setcounter{table}{0}
\makeatletter
\renewcommand{\theequation}{S\arabic{equation}}
\renewcommand{\thefigure}{S\arabic{figure}}

\section{Energy dependence of the sterile–muon system}

In this Supplemental Material we illustrate the energy dependence of the effective
mixing parameters and conversion probabilities in the two scenarios considered in
the main text: (i) the resonant scenario with baryonic sterile neutrino, and (ii) the
non-resonant realization with non-standard interactions. Throughout, we work in the effective two-flavor basis $(\nu_\mu,\nu_s)$.

\begin{figure}[b!]
	\centering
	\includegraphics[width=0.9\textwidth]{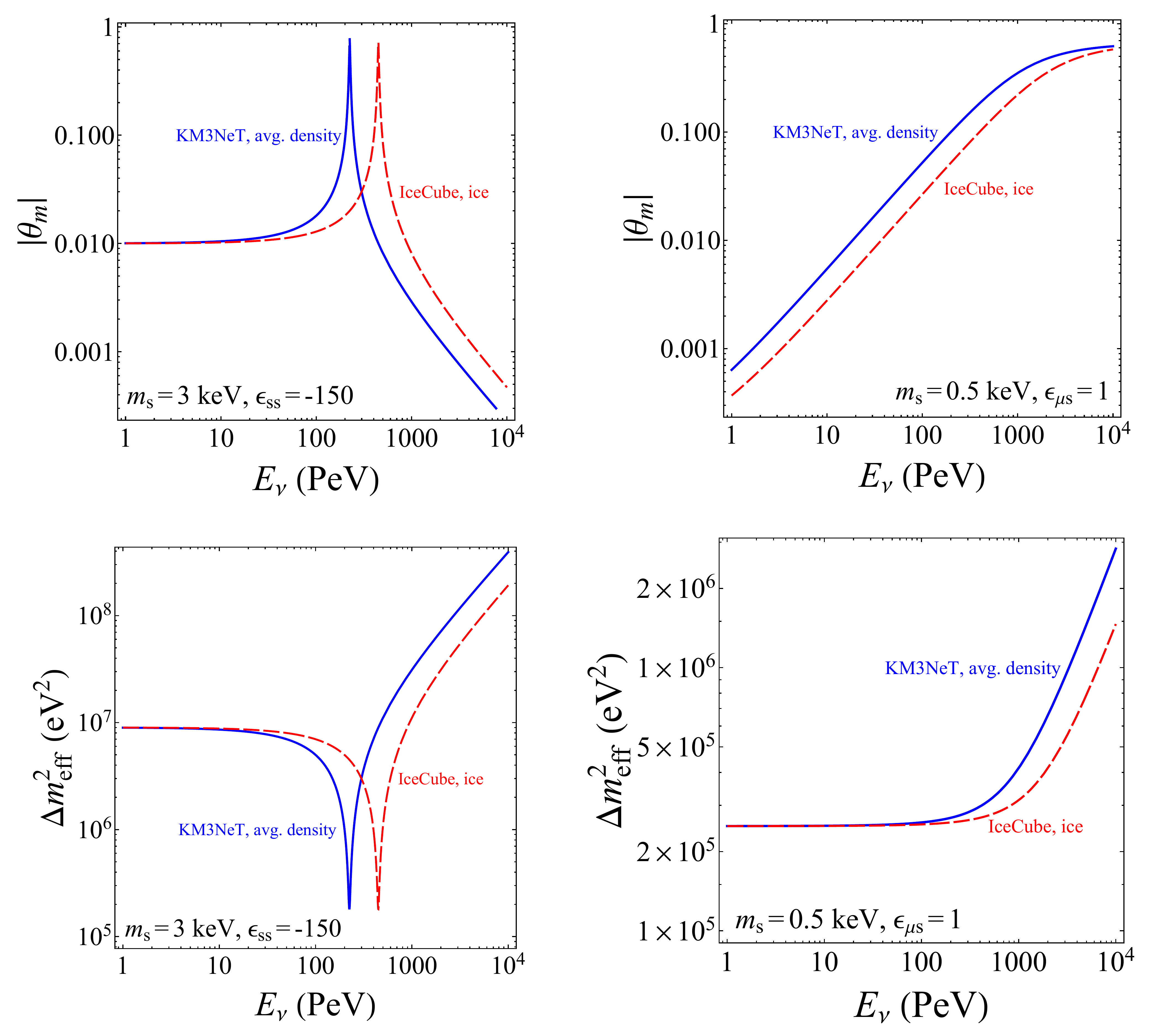}
	\caption{
		Energy dependence of the effective mixing parameters, for the benchmark points shown in the figure.
		\textit{Upper panels:} Absolute value of the effective mixing angle in matter, $|\theta_m(E_\nu)|$, for both the resonant and non-resonant scenarios. For the resonant case, a clear enhancement appears between
		$E_\nu \sim 150$ and $300~\text{PeV}$, corresponding to the resonance region. For the non-resonant non-standard interaction scenario, the enhancement can be observed for $E_\nu \gtrsim \mathcal{O}(100)$~PeV.
		\textit{Lower panels:} The effective mass-squared difference $|\Delta m_{\rm eff}^2(E_\nu)|$, for both scenarios.
	}
	\label{fig:thetammass_S1}
\end{figure}

For the resonant scenario, the Hamiltonian can be expressed as
\begin{equation}
	H_{\rm res}
	=
	\begin{pmatrix}
		c_\theta & s_\theta \\
		-s_\theta & c_\theta
	\end{pmatrix}
	\begin{pmatrix}
		0 & 0 \\
		0 & m_s^2/ (2E_\nu)
	\end{pmatrix}
	\begin{pmatrix}
		c_\theta & -s_\theta \\
		s_\theta &  c_\theta
	\end{pmatrix}
	+
	\begin{pmatrix}
		V_{\rm NC} & 0 \\
		0 & V_s
	\end{pmatrix} \; ,
\end{equation}
where we define $c_\theta \equiv \cos\theta$ and
$s_\theta \equiv \sin\theta$; the sterile neutrino potential is given by $V_s=\sqrt{2} G_F \epsilon_{ss} (n_n+n_p)$, where $n_n$ ($n_p$) is the neutron (proton) number density in matter. In the simplified limit of $n_n = n_p$, the sterile potential takes the form of $V_s \equiv 2 \epsilon_{ss} V_{\rm CC}$. Further, in this effective 2-flavor approach, we assume that $m_s^2 \gg \Delta m^2_{31}$, i.e., any effects of the atmospheric mass-squared difference is negligible.
In the matter basis, in the limit of $|V_s| \gg |V_{\rm NC}|$, the effective mixing angle and mass-squared difference are given by:
\begin{align}
	\theta_m(E_\nu)
	&\approx
	\frac{1}{2}
	\tan^{-1}
	\left[
	\frac{m_s^2 \sin 2\theta /(2E_\nu)}{
		m_s^2 \cos 2\theta /(2E_\nu) + V_s}
	\right] \; ,
	\\[3mm]
	\Delta m_{\rm eff}^2(E_\nu)
	&\approx
	\sqrt{
		m_s^4 + 4 m_s^2 E_\nu V_s \cos 2\theta + 4 E_\nu^2 V_s^2 } \; .
\end{align}

In the non-resonant scenario with non-standard interactions, the vacuum part is analogous to the above one, but the matter term contains an off-diagonal potential term, connecting the $\nu_\mu$ sector to the $\nu_s$ sector:
\begin{equation}
	H_{\rm NSI}
	=
	\begin{pmatrix}
		c_\theta & s_\theta \\
		-s_\theta & c_\theta
	\end{pmatrix}
	\begin{pmatrix}
		0 & 0 \\
		0 & m_s^2/ (2E_\nu)
	\end{pmatrix}
	\begin{pmatrix}
		c_\theta & -s_\theta \\
		s_\theta &  c_\theta
	\end{pmatrix}
	+
	\begin{pmatrix}
		V_{\rm NC} & \epsilon_{\mu s} V_{\rm CC} \\
		\epsilon_{\mu s} V_{\rm CC} & 0
	\end{pmatrix}.
\end{equation}
In the limit of $V_{\rm NC} = - V_{\rm CC}/2$ (i.e., for equal proton and neutron density in matter, $n_n = n_p$), small $\theta$, and large $m_s$, the effective parameters are
\begin{align}
	\theta_m(E_\nu)
	&=
	\frac{1}{2}
	\tan^{-1}
	\left[
	\frac{4\,\epsilon_{\mu s} E_\nu V_{\rm CC}}{m_s^2 + E_\nu V_{\rm CC}}
	\right] \; ,
	\\[3mm]
	m_{\rm eff}^2(E_\nu)
	&=
	\sqrt{(E_\nu V_{\rm CC} + m_s^2)^2
		+ 16\,\epsilon_{\mu s}^2  E_\nu^2 V_{\rm CC}^2 }\; .
\end{align}
For ultra-high energies, once the condition $V_{\rm CC} \gg m_s^2/E_\nu$ is satisfied, we further obtain
$\theta_m \approx \tfrac{1}{2}\tan^{-1}(4\epsilon_{\mu s})$.

\begin{figure}[h!]
	\centering
	\includegraphics[width=0.48\textwidth]{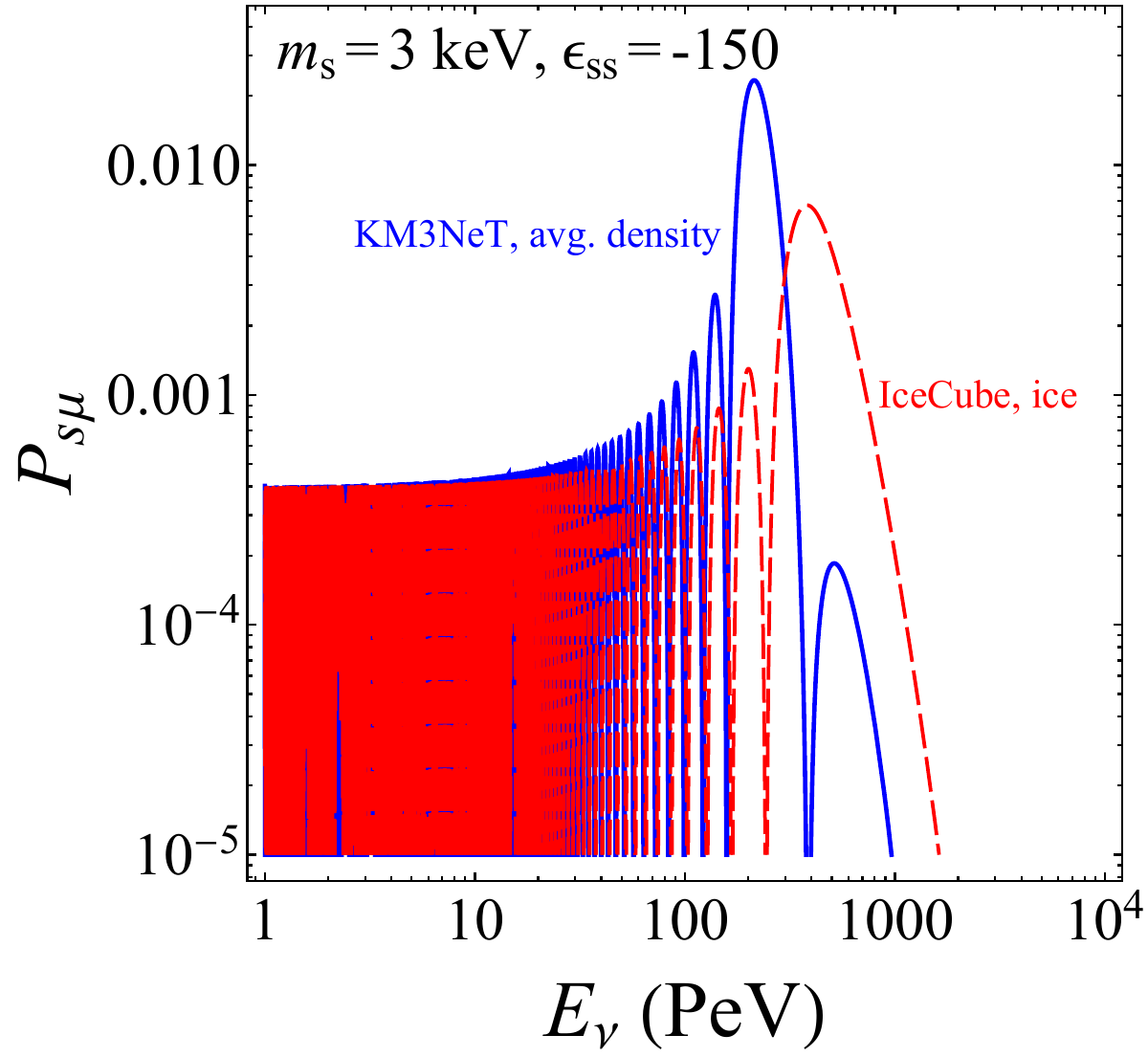}
	\hfill
	\includegraphics[width=0.48\textwidth]{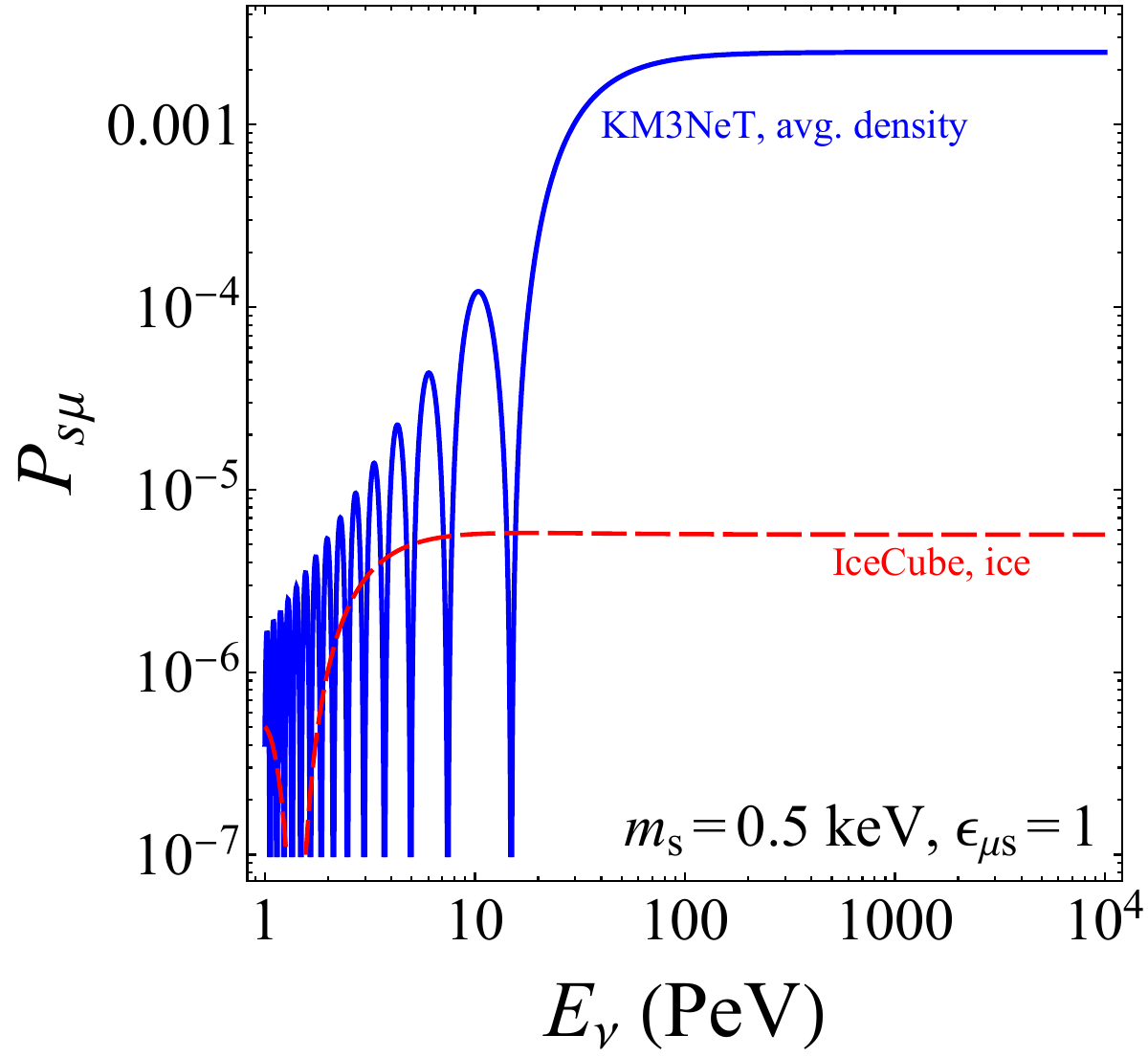}
	\caption{
		Sterile-to-muon neutrino conversion probability, $P_{s\mu}$, as a function of energy, for KM3NeT and IceCube. In the left panel, we present the conversion probability for the resonant scenario, for a benchmark point $m_s = 3$~keV, and $\epsilon_{ss} = -150$, showing a pronounced enhancement of $P_{s\mu}$ in the $E_\nu \sim 150$--$300~\text{PeV}$ range. In the right panel, we show the probability for the non-resonant scenario, for a benchmark of $m_s = 0.5$~keV, and $\epsilon_{\mu s} = 1$, where $P_{s\mu}$ becomes large for $E_\nu \gtrsim \mathcal{O}(100)~\text{PeV}$ without affecting the low-energy regime.
	}
	\label{fig:Psmu_S2}
\end{figure}

Figure~S1 shows $|\theta_m(E_\nu)|$ and $|\Delta m_{\rm eff}^2(E_\nu)|$ for  the representative benchmark points taken for the two scenarios.
For the resonant baryonic sterile neutrino scenario (left panel), we take $m_s = 3$~keV, and $\epsilon_{ss} = -150$.
The resonance condition $2E_\nu V_s = - \, m_s^2 \cos 2\theta$ is satisfied near	$E_\nu \simeq 220~\text{PeV}$, leading to a pronounced peak in $|\theta_m(E_\nu)|$ between $\sim 150$~PeV and $300$~PeV.
In the non-resonant realization (right panel), we take $m_s = 0.5$~keV, and $\epsilon_{\mu s} = 1$. This leads to increase of $|\theta_m(E_\nu)|$ with $E_\nu$.

For a given baseline $L$, the sterile-to-muon neutrino conversion probability in the
two-flavor approximation is
\begin{equation}
	P_{s\mu}
	\;=\;
	\sin^2 2\theta_m(E_\nu)\,
	\sin^2\!\left[
	\frac{\Delta m_{\rm eff}^2(E_\nu)\,L}{4E_\nu}
	\right].
\end{equation}
In Fig.~S2 we show $P_{s\mu}(E_\nu)$ for both the resonant and non-resonant scenarios, with baselines of $147$~km for KM3NeT and $14$~km for IceCube.
For the resonant scenario, the probability $P_{s\mu}$ shows a strong peak in the $E_\nu \sim 150$--$300~\text{PeV}$ interval, where the matter mixing is resonantly enhanced. At significantly lower energies, $P_{s\mu}$ is strongly	suppressed and essentially no low-energy active flux is generated from the	sterile component.
For the non-resonant scenario, $P_{s\mu}$ rises with energy and remains sizable for all $E_\nu \gtrsim \mathcal{O}(100)~\text{PeV}$, while staying small at lower energies.
Both the resonant and non-resonant realizations are therefore constructed such that the low-energy behavior of standard neutrinos is essentially unchanged, while allowing for efficient $\nu_s \to \nu_\mu$ conversion over a broad ultra-high-energy range relevant for the KM3-230213A event.

\end{document}